\documentclass[a4paper,11pt]{article}
\usepackage{pos}

\usepackage{bm}

\newcommand{\nn}{\nonumber}
\renewcommand{\(}{\left(}
\renewcommand{\)}{\right)}

\title{Extraction of genuine twist-3 distribution from data}

\author*[a]{Alexey Vladimirov}

\affiliation[a]{Departamento de Física Teórica \& IPARCOS, Universidad Complutense de Madrid, \\ Plaza de las Ciencias 1, 28040 Madrid, Spain}

\emailAdd{alexeyvl@ucm.es}

\abstract{I report on an extraction of genuine twist-three parton distributions, obtained from a joint fit of the deep-inelastic scattering (DIS) structure function $g_2$, its moment $d_2$, and the Sivers and worm-gear-T asymmetries measured in semi-inclusive DIS (SIDIS). These four observables probe different sections of the same twist-three distributions, and thus their joint analysis makes a precision extraction of this two-variable function possible. Considering a sub-set of observables does not constrain the twist-three distribution in all its parts. Nonetheless, such consideration places a sufficient amount of restrictions to sufficiently accurately estimate eliminated observables, explicitly demonstrating the universality of twist-three distributions. The report is based on refs.\cite{Vladimirov:2025qrh} and \cite{Portela:2026wwn}.}

\FullConference{The 33rd International Workshop on Deep Inelastic Scattering and Related Subjects (DIS2026)\\
4 - 8 May 2026\\
Bologna, Italy\\}

\begin{document}
\maketitle

\section{Introduction}

Parton distribution functions (PDFs) carry the information about the internal structure of the proton. Most phenomenological effort has concentrated on leading-twist densities -- unpolarized, helicity and transversity PDFs -- since they dominate cross-sections. However, these PDFs essentially are the number densities, and carry little information about the dynamics of partons within the proton; to explore this dynamics, one should consider more involved parton distributions, such as transverse momentum distributions \cite{Boussarie:2023izj} (TMDs), generalized parton distributions (GPDs) \cite{Diehl:2003ny} or higher-twist distributions \cite{Braun:2022gzl}. These distributions have rich phenomenological implications, but at the same time they are much harder to determine from the data. All of them possess dependence on several variables, and the corresponding data is generally much noisier.

This report is dedicated to the determination of genuine twist-three distributions made in refs. \cite{Vladimirov:2025qrh, Portela:2026wwn}. Genuine twist-three distributions occupy a special role among higher-twist distributions. Their geometrical twist is preserved by renormalization, which guarantees that they are universal objects \cite{Braun:2009mi, Bukhvostov:1985rn, Balitsky:1987bk}. Being defined by three-parton (quark-gluon-quark or three-gluon) light-cone correlators, they serve a fundamental basis for construction of any other (dynamical) twist-three function. Interest in them dates back to the early eighties, but essentially no phenomenological extraction existed until recent proof-of-concept study \cite{Vladimirov:2025qrh}. The main reason is that a twist-three PDF is a function of two momentum fractions $x_1$ and $x_2$, and any single observable only accesses it via a one-dimensional section or integral convolution. The only realistic strategy is therefore a joint, evolution-consistent fit of several processes, which restricts various parts of the two-dimensional function while the evolution equation adds further correlations to this structure. For this study we selected: the DIS structure function $g_2$, its moment $d_2$, and the Sivers and worm-gear-T (Kotzinian-Mulders) SIDIS asymmetries, as the simplest sources of twist-three data. The analysis is made by the twist-three evolution code \texttt{snowflake} \cite{Rodini:2024usc} merged with the TMD library \texttt{artemide} \cite{artemide, Moos:2023yfa, Moos:2025sal}, and utilizes the methodological frameworks established by our group in the TMD fits \cite{Scimemi:2019cmh, Vladimirov:2019bfa, Bertone:2019nxa, Bury:2020vhj, Bury:2021sue, Horstmann:2022xkk}.

This contribution has a narrow scope. It presents the brief idea of joint data analyses with respect to genuine twist-three PDFs, and some technical elements. The special emphasis is made on the cross-talk between the DIS and TMD sectors of the fit, which is the most convincing phenomenological evidence for the universality of twist-three PDFs and QCD factorization theorems. All other aspects (actual values of twist-three PDFs and TMDs, sum rules, inter-quark forces, nucleon tomography, comparisons with other groups and many others) can be found in \cite{Vladimirov:2025qrh, Portela:2026wwn}, while the code grids and collection of plots are publicly accessible in the repository \cite{REP}.

\section{Twist-three distributions}
\label{sec:tw3-theory}

There are two chiral-even genuine twist-three quark-gluon-quark distributions $T_f$ and $\Delta T_f$, defined as \cite{Braun:2009mi, Rein:2022odl, Rodini:2024usc}
\begin{eqnarray}&&\label{def:T}
\langle p,S |g\,\overline{q}(z_{1}n)F^{\mu+}(z_{2}n)\gamma^{+}q(z_{3}n)|p,S\rangle
\\\nn &&\qquad
=2\epsilon^{\mu\nu}_{T}s_{\nu}(p^{+})^{2}M\int[dx]\:T_f(x_1,x_2,x_3)\:e^{-ip^{+}(x_1z_1+x_2z_2+x_3z_3)}~,
\end{eqnarray}
\begin{eqnarray}
\label{def:DeltaT} &&
\langle p,S |ig\,\overline{q}(z_{1}n)F^{\mu+}(z_{2}n)\gamma^{+}\gamma^{5}q(z_{3}n)|p,S\rangle
\\\nn &&\qquad
=-2s^{\mu}_{T}(p^{+})^{2}M\int[dx]\:\Delta T_f(x_1,x_2,x_3)\:e^{-ip^{+}(x_1z_1+x_2z_2+x_3z_3)}
~,
\end{eqnarray}
with $q$ a quark field of flavor $f$, $F^{\mu\nu}$ the gluon field strength, and $M$ the proton mass; the standard notation for light-cone components is used throughout. The gauge links that make the operators gauge invariant are omitted for brevity. The momentum fractions $x_{1,2,3}$ live on the hexagonal domain $-1\leq x_i\leq1$, $x_1+x_2+x_3=0$, such that $T_f,\Delta T_f$ are effectively functions of two variables. An analogous pair $T_{3F}^\pm$ is defined for the three-gluon correlator $\sim \langle F^{\alpha+}F^{\beta +}F^{\gamma+}\rangle$ \cite{Braun:2009mi, Scimemi:2018mmi, Alvaro:2026nip}. Together these functions $(T,\Delta T,T_{3F}^+,T_{3F}^-)$ form a complete basis, which does not mix with other genuine twist-three distributions under the evolution.

Furthermore, for the solution of the evolution equation, it is convenient to use C-parity-definite combinations $(\mathfrak{S}^\pm_f, \mathfrak{F}^\pm)$ \cite{Braun:2009mi}, which partially diagonalize the evolution equations. Namely, they split into pure ``plus'' and ``minus'' parts, within which, however, quark and gluon distributions mix non-trivially
\begin{eqnarray}\label{def:ev-singlet}
\mu^2 \frac{\partial}{\partial \mu^2} \(\begin{array}{c}
\mathfrak{S}^\pm_{\text{S}}\\
\mathfrak{F}^\pm
\end{array}\)=
-a_s(\mu) \(
\begin{array}{cc}
\mathbb{H}_{qq}     &  \mathbb{H}_{qg} \\
\mathbb{H}_{gq}     &  \mathbb{H}_{gg}
\end{array}\)
\otimes
\(\begin{array}{c}
\mathfrak{S}^\pm_{\text{S}}\\
\mathfrak{F}^\pm
\end{array}\),
\end{eqnarray}
Here $\otimes$ is a 2D convolution over the hexagon, and the kernels $\mathbb{H}$ can be found in refs.~\cite{Braun:2009mi,Rodini:2024usc}. More details on the structure and features of these equations and their numerical implementation can be found in \cite{Rodini:2024usc}. 

Twist-three PDFs possess a rich structure of discrete symmetries, e.g. $T_f(x_{123})=T_f(-x_{321})$, $\Delta T_f(x_{123})=-\Delta T_f(-x_{321})$. These symmetries must be respected within the fitting procedure, and significantly restrict the parametric freedom. In our work, as it is the first extraction of these PDFs, we utilized the simplest ansatz that obeys the symmetries, vanishes at $\|x\|=\max(|x_1|,|x_2|,|x_3|)\to1$, and is smooth except possibly at $\|x\|=0$. The large and small-$\|x\|$ behaviour is common to all flavors, and is governed by
\begin{equation}\label{ansatz1}
h(x_{123}) = \frac{(1-x_1^2)^{a_1} (1-x_2^2)^{a_2} (1-x_3^2)^{a_3}}{(x_1^2 + x_2^2 + x_3^2)^{a_0}}\,,
\end{equation}
where $a_0$ controls the behaviour at $\|x\|\to0$, and $a_{1,2,3}\!>\!0$ control the endpoints. The flavor-dependent part (for $u,d,s$ flavors) is given by the most general quadratic polynomial compatible with the symmetry restrictions
\begin{eqnarray}\label{ansatz2}
\mathfrak{S}_f^+(x_{123};1\text{GeV})&=&h(x_{123}) \(\alpha_0^f+\alpha_{11}^f x_1^2+\alpha_{13}^fx_1x_3+\alpha_{33}^fx_3^2\),
\\
\label{ansatz3}
\mathfrak{S}_f^-(x_{123};1\text{GeV})&=&h(x_{123}) \(\alpha_1^fx_1+\alpha_3^f x_3\)\,.
\end{eqnarray}
The gluon distributions $T_{3F}^\pm$ get a similar, minimal two-parameter ansatz, entering the fit only through the mixing in eq.~(\ref{def:ev-singlet}), since no data set is directly sensitive to the gluon. Also, the non-perturbative part of the Sivers/worm-gear-T TMD distributions (sec.~\ref{sec:cross-talk}) is parametrized by one more, flavor-independent function $f_{\text{NP}}(x,b)=1/\cosh(\lambda|b|)$ for the transverse-momentum profile, with $\lambda=0.5$~GeV fixed, since data cannot constrain $\lambda$ and the ansatz simultaneously \cite{Bury:2021sue, Horstmann:2022xkk}.

This ansatz makes the boundary condition for the evolution-equation system at $\mu_i=1$~GeV. This low scale is selected well below the data, so the main structure and internal consistency of twist-three distributions are generated dynamically via evolution equations at the point of data comparison. Altogether the ansatz has $24$ free parameters, richer than in \cite{Vladimirov:2025qrh}. The results of both fits agree within uncertainties.

\section{Extraction of twist-three distributions}
\label{sec:extraction}

The four processes access the twist-three PDFs through very different integrals, which is what makes their combination informative. The polarized DIS structure function $g_2=g_2^{\text{WW}}+\overline{g}_2$ splits into a Wandzura-Wilczek term $g_2^{\text{WW}}$, fixed by the helicity PDF \cite{Wandzura:1977qf}, and a genuine twist-three piece $\overline{g}_2$, a 2D integral of $\mathfrak{S}^+$. Its lowest moment $d_2\propto\int[dx]\,T_f$ normalizes $T_f$ over the whole hexagon; it is measured semi-independently, and is also accessible via lattice QCD. In SIDIS, the single-spin asymmetry $A_{UT}^{\sin(\phi_h-\phi_S)}$ and double-spin asymmetry $A_{LT}^{\cos(\phi_h-\phi_S)}$ are TMD convolutions of the Sivers ($f_{1T}^\perp$) and worm-gear-T ($g_{1T}^\perp$) distributions with the unpolarized FF. At small $b$ these match onto $T$ and $\Delta T$ \cite{Scimemi:2018mmi, Moos:2020wvd} (sec.~\ref{sec:cross-talk}), $f_{1T}^\perp$ reducing to $T$ on the line $x_2=0$ (Qiu-Sterman function \cite{Qiu:1991pp,Efremov:1983eb}). Together, $\{d_2,\overline{g}_2,A_{UT}^{\sin(\phi_h-\phi_S)},A_{LT}^{\cos(\phi_h-\phi_S)}\}$ project $T,\Delta T,T_{3F}^\pm$ onto different lines/integrals of the hexagon, and, correlated by evolution, allow a 2D reconstruction from 1D measurements.

There are many experiments that provide data for these observables. The complete review is given in \cite{Portela:2026wwn}. Unfortunately, most of the data is either taken at very low energy or has inadequately large uncertainties. So, selecting the most modern and precise experiments, and applying the cut $Q^2>2$~GeV$^2$ and $p_{h\perp}^2/z^2Q^2<0.35$ (for SIDIS measurements), we found 372 data points in total ($168$ for $g_2$ by E143/E154/E155, HERMES, JLab Hall A; $13$ for $d_2$ by E143/E155, HERMES, JLab Hall A, SANE, and a lattice measurement by RQCD \cite{Burger:2021knd}; $117$ for $A_{UT}^{\sin(\phi_h-\phi_S)}$ and $74$ for $A_{LT}^{\cos(\phi_h-\phi_S)}$ by COMPASS, HERMES, JLab).

One of the central and most expansive parts of the analysis is the propagation of uncertainties. We identify two sources: the data uncertainty and the PDF uncertainty (which is known to be the dominant theoretical uncertainty especially for TMD extractions \cite{Bury:2022czx}). Uncertainties are propagated with a parametric bootstrap, as in our earlier TMD extractions \cite{Bury:2021sue, Moos:2025sal}, which is in turn inherited from collinear PDF fits \cite{Ball:2008by}. For each of $300$ replicas we randomize: (i) the external input (helicity PDF \cite{Bertone:2024taw}, unpolarized TMD PDF/FF \cite{Moos:2025sal}), and (ii) the data, via pseudo-data generated from the reported correlated/uncorrelated uncertainties. Furthermore, minimization of each replica starts from a random point to avoid bias. Minimization uses a $\chi^2$ weighted by $1/N_k$ over the four subsets $k=\{d_2,g_2,UT,LT\}$, so that none of them (especially the large $g_2$ set) dominates the joint fit.

The fit converges to a well-defined minimum, with
\begin{eqnarray}\label{chi2:main}
\frac{\chi^2_{d2}}{N_{d2}}=1.01^{+0.12}_{-0.13},\quad
\frac{\chi^2_{g2}}{N_{g2}}=0.99^{+0.05}_{-0.05},\quad
\frac{\chi^2_{UT}}{N_{UT}}=1.06^{+0.05}_{-0.05},\quad
\frac{\chi^2_{LT}}{N_{LT}}=0.94^{+0.03}_{-0.03},
\end{eqnarray}
and total $\chi^2_{\text{tot}}/N_{\text{tot}}=1.01^{+0.03}_{-0.03}$: all four observables are described simultaneously by the same $24$-parameter ansatz. Comparison with the null hypothesis confirms observation of a genuine twist-three signal at the $2$--$3\sigma$ level. The detailed discussion on this point is presented in ref.~\cite{Portela:2026wwn}.

\begin{figure}[t]
\centering
\includegraphics[width=0.97\textwidth]{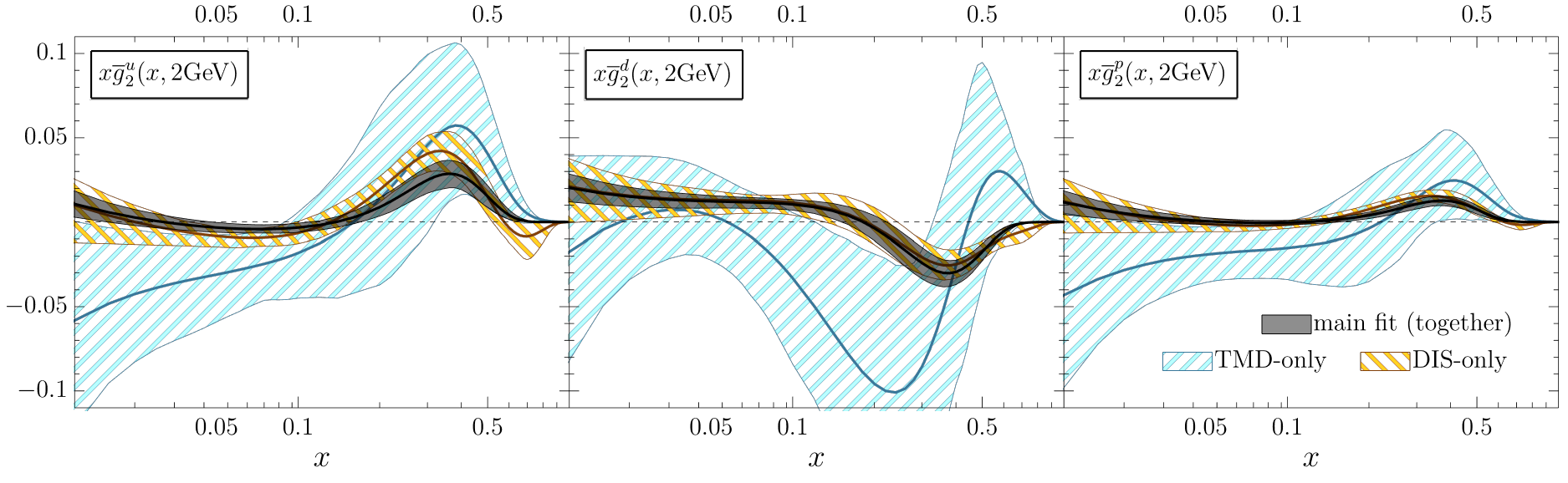}

\vspace{2pt}
\includegraphics[width=0.97\textwidth]{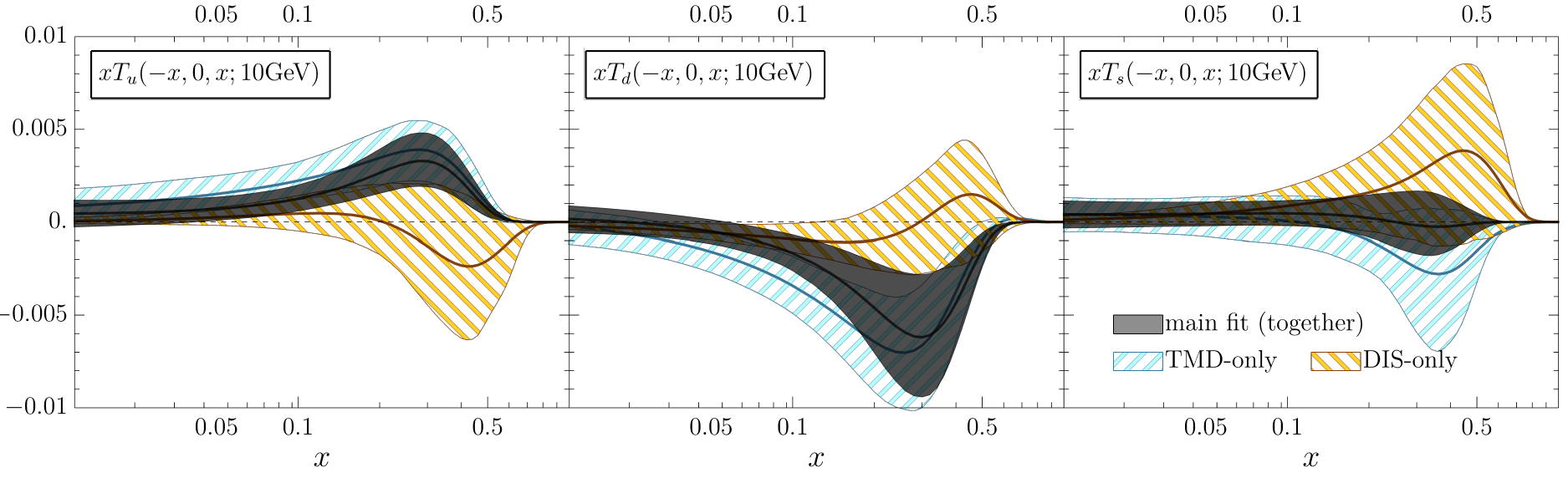}

\vspace{2pt}
\includegraphics[width=0.97\textwidth]{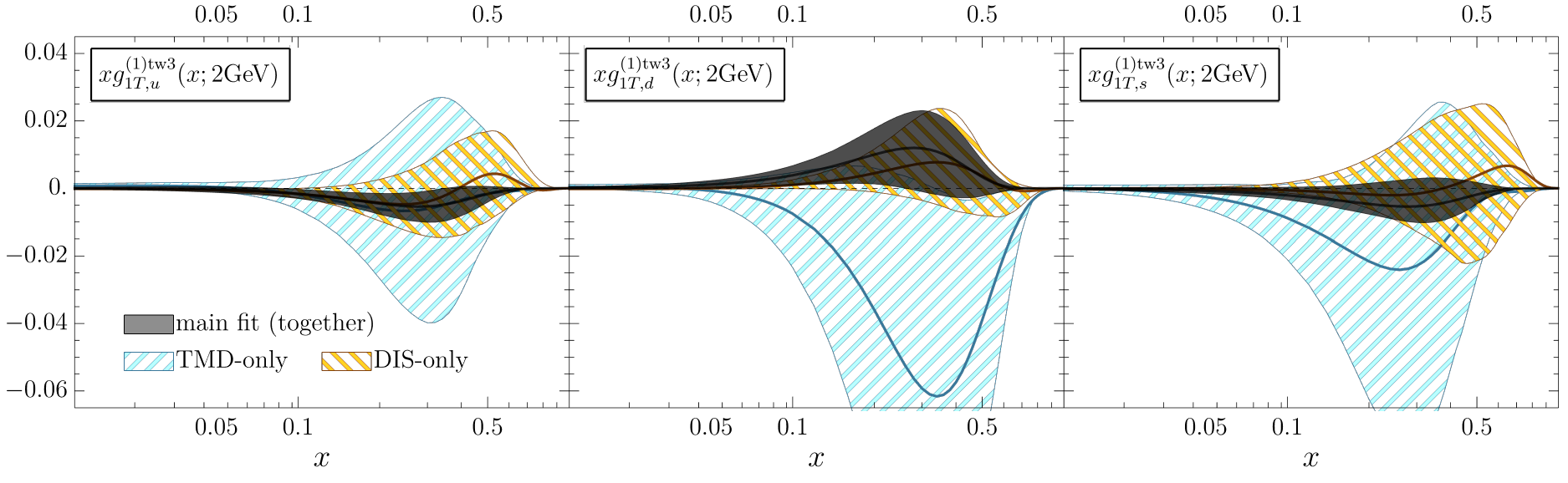}
\caption{\label{fig:crosstalk} Reconstruction of $\overline{g}_2$ (top), of the Qiu-Sterman function $T(-x,0,x)$ (middle), and of the first moment of the twist-three part of the worm-gear-T function (bottom), from the main fit and from the DIS-only and TMD-only fits. The pictures are taken from ref.~\cite{Portela:2026wwn}.}
\end{figure}

\section{Cross-talk between DIS and TMD}
\label{sec:cross-talk}

A central feature of this analysis is that DIS and TMD/SIDIS observables are not independent probes of unrelated objects -- they are different projections of the same genuine twist-three PDFs, in both directions. At small transverse distance $b$, TMD distributions are matched via an OPE onto collinear ones. For instance, the Sivers function, at leading order \cite{Scimemi:2019gge,Scimemi:2018mmi,Rein:2022odl}, reads
\begin{eqnarray}\label{TMD:sivers-smallb}
f_{1T,q}^\perp(x,b)=-\pi\, T(-x,0,x;\mu_{\text{OPE}})+\mathcal{O}(\alpha_s)+\mathcal{O}(b^2)\,.
\end{eqnarray}
In this way, the Sivers function directly measures the twist-three PDF $T$ on the line $x_2=0$ (Qiu-Sterman function). An analogous, although more involved, relation expresses $g_{1T}^\perp$ in terms of $T$ and $\Delta T$. So SIDIS constrains the same functions that enter $g_2$ and $d_2$, and thus, by analyzing SIDIS asymmetries, one can predict DIS measurables. The converse also holds: fixing $T$ and $\Delta T$ from DIS leads to a prediction of the small-$b$ Sivers/worm-gear-T TMDs with no extra parameter. This is what makes it meaningful to speak of \emph{one} set of genuine twist-three distributions, rather than four unrelated fitted shapes. 

To test this quantitatively, we have performed a \emph{DIS-only} fit ($\overline{g}_2,d_2$ with $181$ pt., $\mathfrak{S}^-,T_{3F}^-$ set to zero) and a \emph{TMD-only} fit ($A_{UT}^{\sin(\phi_h-\phi_S)},A_{LT}^{\cos(\phi_h-\phi_S)}$ with $191$ pt.). The central values of the $24$ parameters from both fits are compatible with each other and with the main fit within uncertainties (only $2$ of $24$ parameters differ at the $1\sigma$ level). This means that DIS and TMD data prefer the \emph{same} distributions, constraining different combinations of parameters with different precision. From the perspective of the $\chi^2$ distribution of replicas, each fit describes well only its own data ($\chi^2/N\approx1$), but grossly fails on the other's. Particularly, the DIS-only fit gives $\chi^2_{UT}/N_{UT}=8.2$, $\chi^2_{LT}/N_{LT}=1.8$, and the TMD-only fit gives $\chi^2_{d2}/N_{d2}\approx47$, $\chi^2_{g2}/N_{g2}\approx32$. It indicates that the uncertainties of the distributions are too large in a ``distinct'' sector, although the average value is similar. So combining them tightens correlations without shifting the central shape, which gives the first direct phenomenological demonstration of twist-three universality across the DIS/TMD divide.

Figure~\ref{fig:crosstalk} demonstrates the comparison of shapes of distributions obtained in different fits, confirming the cross-talk at the level of the reconstructed functions themselves. Evidently, the agreement for $\overline{g}_2$ and $g_{1T}^{(1),\text{tw3}}$ (fig.~\ref{fig:crosstalk}, top \& bottom) is very good. This is because these functions share the same domain of integration, although with different evolution kernels. It is interesting that the DIS-only extraction of the $g_{1T}^{(1),\text{tw3}}$ TMD gives even better precision than the TMD-only fit, which is due to more precise data in the DIS measurements. The Qiu-Sterman function (middle) shows a moderate agreement, since TMD-only data directly constrains only $T(-x,0,x)$ (\ref{TMD:sivers-smallb}) itself, while $\overline{g}_2,d_2$ are sensitive to derivatives/integrals of $T$ along other directions. Thus, the DIS-only reconstruction of this line is a genuinely indirect, evolution-driven prediction, with correspondingly larger uncertainty. Even so, different lines agree with each other within uncertainties.

\section{Conclusion}

The present analysis clearly demonstrates that the future of the multi-dimensional parton studies lies in the joint analysis of TMD and collinear distributions. These data allow us to restrict parts of distributions unreachable by other experiments. Even if the data for some class of observables is worse, it still brings novel information. For twist-two observables, the effect is mild, since DIS data is significantly more precise than TMD data. Nonetheless, SIDIS data brings finer sensitivity to the flavor decomposition \cite{Barry:2025glq} for unpolarized PDFs. For the case of twist-three distributions, the joint consideration is fundamental and allows us, for the first time, to observe genuine twist-three distributions. The present analysis already combines four processes, but the list of twist-three-sensitive observables (other SIDIS/Drell-Yan asymmetries, jets, lattice matrix elements) is considerably longer, and each one added is expected to further pin down these distributions and test universality anew -- the natural direction for future work.

\smallskip
\noindent\textit{Acknowledgments.} The author would like to thank the Spanish Ministerio de Ciencia e Innovaci\'on (grant No. PID2022-136510NB-C31 funded by MCIN/AEI/10.13039/501100011033) for the support of this project and for participation in the conference. The project is also funded by the Atracci\'on de Talento Investigador program of the Comunidad de Madrid (Spain) No. 2020-T1/TIC-20204.

\bibliographystyle{JHEP}
\bibliography{bibFILE}

\end{document}